\documentclass[conference]{IEEEtran}

\IEEEoverridecommandlockouts
\usepackage{cite}
\usepackage{amsmath,amssymb,amsfonts}
\usepackage{algorithmic}
\usepackage{graphicx}
\usepackage{textcomp}
\usepackage{xcolor}
\usepackage{dsfont}

\def\BibTeX{{\rm B\kern-.05em{\sc i\kern-.025em b}\kern-.08em
    T\kern-.1667em\lower.7ex\hbox{E}\kern-.125emX}}

\usepackage{mathtools}

\usepackage{steinmetz}

\usepackage[ruled,norelsize]{algorithm2e}
\makeatletter
\newcommand{\removelatexerror}{\let\@latex@error\@gobble}
\makeatother
\usepackage{caption}
\usepackage{subcaption}

\begin{document}

\title{Rate-Splitting Multiple Access for Semantic-Aware Networks: an Age of Incorrect Information Perspective}



\author{Onur~Dizdar,~\IEEEmembership{Member,~IEEE,}
        and~Stephen~Wang,~\IEEEmembership{Senior~Member,~IEEE}
\thanks{The authors are with VIAVI Marconi Labs, VIAVI Solutions Inc., Stevenage SG12AN, UK. (e-mail: onur.dizdar@viavisolutions.com; stephen.wang@viavisolutions.com.)}
\thanks{This work is supported in part by the UK Department for Science, Innovation and Technology under the Future Open Networks Research Challenge project TUDOR (Towards Ubiquitous 3D Open Resilient Network). The views expressed are those of the authors and do not necessarily represent the project.}}

\maketitle

\begin{abstract}
In this letter, we design a downlink multi-user communication framework based on Rate-Splitting Multiple Access (RSMA) for semantic-aware networks. First, we formulate an optimization problem to obtain the optimal user scheduling, precoding, and power allocation schemes jointly. We consider the metric Age of Incorrect Information (AoII) in the objective function of the formulated problem to maximize the freshness of the overall information to be transmitted. Using big-M and Successive Convex Approximation (SCA) methods, we convert the resulting non-convex problem with conditional objective and constraints into a convex one and propose an iterative algorithm to solve it. By numerical results, we show that RSMA achieves a lower AoII than SDMA owing to its superior performance under multi-user interference. 
\end{abstract}


\IEEEpeerreviewmaketitle

\section{Introduction}

Rate-Splitting Multiple Access (RSMA) is a multiple access (MA) technique for multi-antenna systems that relies on rate-splitting at the transmitter and successive interference cancellation (SIC) at the receivers. 
RSMA manages multi-user interference by partially decoding interference and partially treating it as noise, and has been shown to outperform existing MA schemes, such as Space Division Multiple Access (SDMA) and Non-Orthogonal Multiple Access (NOMA) \cite{mao_2022, dizdar_2020, mao_2018, mishra_2022}.

Although its roots can be traced back to the seminal work of Weaver \cite{weaver_1949}, semantic communications has become another recently emerging concept to address the requirements of next generation networks with the advancement of machine learning applications. 
Semantic communications focuses on the communication problem beyond the ``Technical Level'', and aims to convey meaning or features of  information instead of transmitting correct bits.    
It is expected to change the way we design and measure the performance of wireless networks, since the conventional metrics fail to capture its potential and waste the available resources. Consequently, new metrics have been proposed to capture the needs of semantic networks, such as, Age of Incorrect Information (AoII).
AoII is proposed in \cite{maatouk_2023} to address the shortcomings of Age of Information (AoI) by capturing the effects of content and data freshness together, and has been considered among the candidate metrics to be used for designing semantic-aware networks \cite{uysal_2022, get_2023}. 

There are a few works that study RSMA in the context of semantic communications or AoI-based metrics \cite{yang_2023, akyurek_2023}. In \cite{yang_2023}, the authors consider RSMA for joint communication and computation in semantic communication networks. 
In \cite{akyurek_2023}, the author investigates the AoI performance of RSMA for a two-user single-antenna downlink communication system over an AoI minimization problem to find the optimal power allocation policy. 
The results show that RSMA outperforms other MA techniques to address the needs of semantic-aware networks.

In this paper, we design an RSMA-based scheduling scheme for semantic-aware networks. 
The proposed scheme targets applications where a monitor observes one or more events and aims to keep its users updated about the events, and ensure that the information at its users is fresh content wise, such as video streaming and fire monitoring \cite{maatouk_2023}. Another example is the use of satellites for remote monitoring, such as disaster monitoring, disaster relief, aviation and navigation monitoring, where timely transmission of relevant status updates to multiple terminals is vital, as obsolete or outdated information can result in unwanted results \cite{hong_2023}.
We formulate an AoII minimization problem to find optimal user scheduling, power allocation, and precoder design for RSMA in a multi-antenna network with arbitrary number of users. The formulated problem is non-convex and consists of conditional objective and constraints. 
We use big-M and Successive Convex Approximation (SCA) methods to propose an algorithm to solve the formulated problem. We show that RSMA achieves a lower AoII than SDMA with the proposed algorithm.
This is the first work that designs user scheduling, power allocation, and precoding jointly to minimize AoII for RSMA in a multi-antenna network with arbitrary number of users. 



\section{System Model}
We consider a system where a single transmitter with $N$ antennas performs downlink communications with $K$ single-antenna receivers with the index set $\mathcal{K}=\{1,\ldots, K\}$. We assume that transmission is performed in a time-slotted framework and the time is normalized to the slot duration, so that, the slot duration is taken as $1$ and each transmission occurs at a time instance $t \in \mathbb{N}$. An information process $X_{t,k}$, $k \in \mathcal{K}$, is observed for each user at the transmitter, which can change over time. Each user has an estimate of the process $X_{t,k}$ at time instance $t$, denoted by $\widehat{X}_{t,k}$. The aim of the transmitter is to update the users about the changes in $X_{t,k}$, {\sl i.e.}, keep $\widehat{X}_{t,k}$ up-to-date. Accordingly, the transmitter may decide to transmit to user-$k$ at time instance $t$ if it observes a certain amount of change in $X_{t,k}$. 

If the transmitter decides to send updates at time instance $t$, it schedules a particular number of users for transmission. We consider RSMA for transmission to the scheduled users. We denote with the set of indexes for such users as $\mathcal{L}_{t} \subseteq \mathcal{K}$. The transmission to the scheduled users start at time instance $t$ and finishes at $t+1$. If user-$k$ can successfully receive the update intended for it, it updates its estimate at $t+1$. We consider the model $\widehat{X}_{t,k} = X_{U_{t,k}}$ for the process estimate at user-$k$, where $U_{t,k}$ is the timestamp of the last successfully received packet by the receiver at time $t$. 

If user-$k$ can decode its update successfully, it sends back an acknowledgement (ACK) packet. We assume that the transmission time for ACK to be negligible, which is a widely used assumption in literature \cite{maatouk_2023}. If an ACK is not received at time instance $t+1$ from user-$k$, the transmitter drops the update $X_{t,k}$ and generates a new update $X_{t+1,k}$.

\subsection{Rate-Splitting Multiple Access}
RSMA performs message splitting at the transmitter by splitting user messages into common and private parts. 
Let us denote the message of user-$k$ at time $t$, its common and private parts as $W_{k}$, $W_{c,k}$ and $W_{p,k}$, respectively. We drop the subscript $t$ from the rate expressions for ease of notation. The transmitter combines $W_{c,k}$, $\forall k \in \mathcal{K}$, into the common message $W_{c}$. The common message $W_{c}$ and the private messages $W_{p,k}$ are independently encoded into streams $s_{c}$ and $s_{k}$, respectively. Linear precoding is applied to all streams and the precoders for the common stream and the private stream of user-$k$ are denoted as $\mathbf{p}_{c} \in\mathbb{C}^{N}$ and $\mathbf{p}_{k} \in\mathbb{C}^{N}$, respectively. Accordingly, the transmitted and received signals are expressed as
\begin{align}
	\mathbf{x}=\mathbf{p}_{c}s_{c}+\sum_{k=1}^{K}\mathbf{p}_{k}s_{k}, \quad \mathbf{y}_{k}&=\mathbf{h}_{k}^{H}\mathbf{x}+z_{k}, \quad \forall k \in \mathcal{K},
	\label{eqn:transmit_signal}	
\end{align}
where \mbox{$\mathbf{h}_{k} \in \mathbb{C}^{N}$} is the channel vector and \mbox{$z_{k} \sim \mathcal{CN}(0,\sigma_{n}^{2})$} is the Additive White Gaussian Noise (AWGN) at user-$k$.
We assume that \mbox{$\mathbb{E}\left\lbrace \mathbf{s}\mathbf{s}^{H}\right\rbrace =\mathbf{I}$} for \mbox{$\mathbf{s}=[s_{c}, s_{1}, \ldots, s_{K}]$}. We also assume that the transmitter has perfect channel state information at transmitter (CSIT) to calculate the precoders.

At the receiver side, the messages are detected using SIC. Each user decodes the common stream first to obtain a common message estimate $\widehat{W}_{c}$ and extract its common message estimate $\widehat{W}_{c,k}$ by treating the private streams for all users as noise. Then, the common stream is reconstructed using $\widehat{W}_{c}$ and subtracted from the received signal. Each user decodes their intended private message $\widehat{W}_{p,k}$ by treating the private messages of other users as noise. Finally, each user combines the message estimates $\widehat{W}_{c,k}$ and $\widehat{W}_{p,k}$ to obtain its original message estimate $\widehat{W}_{k}$. 
Assuming perfect channel state information at receiver (CSIR) and SIC, we express the SINRs for the common and private streams at user-$k$ by
\begin{align}
\gamma_{c,k}=\frac{|\mathbf{h}_{k}^{H}\mathbf{p}_{c}|^{2}}{1+\sum_{i \in \mathcal{K}}|\mathbf{h}_{k}^{H}\mathbf{p}_{i}|^{2}}, \
\gamma_{k}=\frac{|\mathbf{h}_{k}^{H}\mathbf{p}_{k}|^{2}}{1+\sum_{\substack{i \in \mathcal{K},\\ i \neq k}}|\mathbf{h}_{k}^{H}\mathbf{p}_{i}|^{2}}, 
\label{eqn:private}
\end{align}
respectively. 
We define the achievable rates $R_{k}=\log_{2}(1+\gamma_{k})$ and $R_{c,k}=\log_{2}(1+\gamma_{c,k})$ bits/s/Hz. Noting that the common message should be decodable by all users, we define its achievable rate as $R_{c}=\min_{k \in \mathcal{L}_{t}}R_{c,k}$.
We also define the common rate portion for user-$k$ as $c_{k}$, such that, $\sum_{k \in \mathcal{L}_{t}}c_{k}=R_{c}$. 


\subsection{Age of Incorrect Information}

Traditional metrics used for designing wireless networks, such as throughput, delay, and block error rate, do not consider packets' content and can lead to performance bottlenecks with increasing traffic demand. AoI has been proposed to address the shortcomings of such metrics by capturing the time elapsed since the last successfully received update at the monitor was generated at the transmitter. Although AoI aims to ensure the updates at the monitor are fresh, it falls short of capturing the information content of the updates and may result in unnecessary penalties even when the monitor has perfect knowledge about the process at the transmitter. AoII is proposed to address the shortcomings of AoI and other metrics by capturing the significance of a data packet in terms of content and timeliness. AoII aims to ensure that the information that a monitor has about a process at the transmitter is up-to-date and assigns no penalty as long as this condition lasts. When the information at the transmitter and the monitor are different enough, a penalty is applied to ensure that a fresh update is received by the monitor \cite{maatouk_2023}. 

Let us define $V_{t,k}$ as the last time instant where user-$k$ had sufficiently accurate information about the process $X_{t,k}$. The AoII is defined as \cite{maatouk_2023}
\begin{align}
	\Delta_{\mathrm{AoII},k}(X_{t,k}, \widehat{X}_{t,k}, t)=f(t) \times g(X_{t,k}, \widehat{X}_{t,k}).
	\label{eqn:aoii}
\end{align}

In \eqref{eqn:aoii}, $f(t)$ increasingly penalizes the system as long as the mismatch between $\widehat{X}_{t,k}$ and $\widehat{X}_{t,k}$ lasts, and $g(X_{t,k}, \widehat{X}_{t,k})$ quantifies the gap between $\widehat{X}_{t,k}$ and $\widehat{X}_{t,k}$ \cite{maatouk_2023}. Several alternatives have been proposed for $f$ and $g$, such as,
$f_{\mathrm{lin}}(t)=t-V_{t,k}$, $f_{\mathrm{thr}}(t)=\mathds{1}\{t-V_{t,k}\geq \zeta \}$, 
and
$g_{\mathrm{sq}}(X_{t,k}, \widehat{X}_{t,k})=(X_{t,k}-\widehat{X}_{t,k})^{2}$, $g_{\mathrm{thr}}(X_{t,k}, \widehat{X}_{t,k})=\mathds{1}\{|X_{t,k}-\widehat{X}_{t,k}|\geq c\}$, 
where $\zeta$ and $c$ are predefined thresholds \cite{maatouk_2023}. The $f$ and $g$ with $\mathds{1}\{.\}$ function are used to design systems where small mismatches are tolerated, while the ones with linear and square functions are used to penalize any mismatch.

\section{Problem Formulation and Proposed Solution}
\label{sec:problem}
In this section, we formulate an AoII minimization problem to jointly optimize user scheduling, power allocation and precoding for an RSMA-based semantic-aware system design and propose an algorithm to solve it.
Let us assume that a sample is taken at time slot $t$ for each user and $g(X_{t,k},\widehat{X}_{t,k})\neq 0$, $\forall k \in \mathcal{K}$. Transmitter aims to schedule the users for transmission at slot $t$ to minimize the AoII at slot $t+1$. 
We assume a slow process for $X_{t,k}$, so that $g(X_{t+1,k},\widehat{X}_{t+1,k})=0$ if the transmission decision is given and it is successful. Otherwise, the transmission decision may fail to improve the freshness of the process estimate ({\sl e.g.}, achieve $\Delta_{\mathrm{AoII},k}(X_{t+1,k}, \widehat{X}_{t+1,k}, t+1)=0$) and even the successfully transmitted packets become obsolete by the time they are received \cite{maatouk_2023}. 

Let us define $I_{k}$
as the rate required to successfully transmit the update $X_{t,k}$ in the interval between $t$ and $t+1$ to achieve $g(X_{t+1,k},\widehat{X}_{t+1,k})=0$. An update is successfully transmitted if $c_{k}+R_{k} \geq I_{k}$. Recalling that $X_{t,k}$ is a slow process, a successful transmission of the update for user-$k$ results in $\Delta_{\mathrm{AoII},k}(X_{t+1,k}, \widehat{X}_{t+1,k}, t+1)=0$.
Then, AoII at $t+1$ is 
\begin{align}
	&\Delta_{\mathrm{AoII},k}(X_{t+1,k}, \widehat{X}_{t+1,k}, t+1)\nonumber \\
	&=\hspace{-0.1cm}
	\begin{cases}
    \hspace{-0.1cm}(t\hspace{-0.1cm}+\hspace{-0.1cm}1\hspace{-0.1cm}-\hspace{-0.1cm}V_{t+1,k})g(X_{t+1,k},\widehat{X}_{t+1,k})\hspace{-0.3cm}&, \text{if } c_{k}+R_{k} \leq I_{k}, \\
    \hspace{2.5cm}0&, \text{otherwise.}
\end{cases}
\label{eqn:aoii_new}
\end{align}
Defining $\mathcal{L}=\{k \in \mathcal{K}:c_{k}+R_{k} \geq I_{k}\}$ as the set of scheduled users and $\mathbf{P}=\left[\mathbf{p}_{c},\mathbf{p}_{1},\ldots,\mathbf{p}_{K}\right] $, we formulate the following optimization problem to obtain the optimal scheduling, precoder, and power allocation scheme.
\begin{subequations}
	\begin{alignat}{3}
	\min_{\mathbf{c}, \mathbf{P}}&     \quad  \sum_{k\in\mathcal{K}}\Delta_{\mathrm{AoII},k}(X_{t+1,k}, \widehat{X}_{t+1,k}, t+1) \label{eqn:obj}  \\
	\text{s.t.}&  \quad  \sum_{k^{\prime} \in \mathcal{K}}c_{k^{\prime}} \leq R_{c,l}(\mathbf{P}), \quad  \forall l \in \mathcal{L}, \label{eqn:common_rate_1} \\
	& \quad   \mathbf{c} \geq \mathbf{0}, \label{eqn:common_rate_2} \\
	& \quad \mathrm{tr}\left( \mathbf{P}\mathbf{P}^{H}\right)\leq P_{total}. \label{eqn:transmit_power}  
	\end{alignat}
	\label{eqn:formulation_1}
\end{subequations}


The objective \eqref{eqn:obj} and constraint \eqref{eqn:common_rate_1} are conditional. We use big-M method to remove these conditions. 
Beyond its use in solving linear programming problems with Simplex method, Big-M can also be used with a binary variable to enforce constraints. In such applications, the multiplication of $M$ and the binary variable is used to force the equality of two variables. Although the name suggests that $M$ should be chosen very large, extra care should be taken while choosing it, since a too large $M$ value leads to convergence problems and suboptimal solutions in solvers, whereas a too small $M$ value cuts of feasible solutions from the search space. 

Let us define $\mathbf{z}=[z_{1}, z_{2}, \ldots, z_{K}]$ with the binary variable $z_{k} \in \{0,1\}$ to denote the case where user-$k$ is not scheduled at time $t$, {\sl i.e.}, $z_{k}=1$ if $k \notin \mathcal{L}$ and $z_{k}=0$ otherwise. 
Then, we convert \eqref{eqn:formulation_1} into the following.
\begin{subequations}
	\begin{alignat}{3}
	&\min_{\mathbf{c}, \mathbf{P}, \mathbf{z}}\ \  \sum_{k\in\mathcal{K}}(t+1-V_{t+1,k})g(X_{t+1,k},\widehat{X}_{t+1,k})z_{k}   \\
	&\ \ \text{s.t.}  \quad I_{k}-c_{k}-R_{k} \leq Mz_{k}, \quad  \forall k \in \mathcal{K}, \label{eqn:bigM1} \\
	&  \quad\quad\quad c_{k}+R_{k}-I_{k} \leq M(1-z_{k}), \quad  \forall k \in \mathcal{K}, \label{eqn:bigM2} \\
	& \quad\quad\quad  \sum_{k^{\prime} \in \mathcal{K}}c_{k^{\prime}} \leq R_{c,k}(\mathbf{P})+Mz_{k}, \quad  \forall k \in \mathcal{K}, \label{eqn:common_rate_3} \\
	& \quad\quad\quad   z_{k} \in \{0,1\}, \quad  \forall k \in \mathcal{K}, \label{eqn:z_1} \\
	& \quad\quad\quad   \eqref{eqn:common_rate_2}, \eqref{eqn:transmit_power}. \nonumber   
	\end{alignat}
	\label{eqn:formulation_2}
\end{subequations}
\hspace{-0.2cm}The constraints \eqref{eqn:bigM1}-\eqref{eqn:z_1} are non-convex, which makes \eqref{eqn:formulation_2} a non-convex problem. We  can relax the binary requirement in \eqref{eqn:z_1} to obtain a convex constraint. The constraints \eqref{eqn:bigM1}-\eqref{eqn:common_rate_3} can be converted into convex ones using SCA. We introduce $\boldsymbol{\alpha}=[\alpha_{1}, \ldots, \alpha_{K}]$, $\boldsymbol{\beta}=[\beta_{1}, \ldots, \beta_{K}]$, $\boldsymbol{\omega}=[\omega_{1}, \ldots, \omega_{K}]$, $\boldsymbol{\sigma}_{p}=[\sigma_{p,1}, \ldots, \sigma_{p,K}]$, $\boldsymbol{\sigma}_{c}=[\sigma_{c,1}, \ldots, \sigma_{c,K}]$ to write 
\begin{subequations}
	\begin{alignat}{3}
		&\min_{\substack{\mathbf{c}, \mathbf{P}, \mathbf{z}, \boldsymbol{\alpha}, \\ \boldsymbol{\beta}, \boldsymbol{\omega}, \boldsymbol{\sigma}_{p}, \boldsymbol{\sigma}_{c}}} \sum_{k\in\mathcal{K}}(t+1-V_{t+1,k})g(X_{t+1,k},\widehat{X}_{t+1,k})z_{k}  \\
		&\  \text{s.t.}  \ I_{k}-c_{k}-\log_{2}(1+\alpha_{k}) \leq Mz_{k}, \quad  \forall k \in \mathcal{K}, \label{eqn:bigM1_prefinal} \\
		&  \quad \ \ c_{k}+\log_{2}(1+\beta_{k})-I_{k} \leq M(1-z_{k}), \  \forall k \in \mathcal{K}, \label{eqn:bigM2_prefinal} \\
		& \quad \  \sum_{k^{\prime} \in \mathcal{K}}c_{k^{\prime}} \leq \log_{2}(1+\omega_{k})+Mz_{k}, \quad  \forall k \in \mathcal{K} \label{eqn:common_rate_1_prefinal} \\
		& \quad\  \frac{|\mathbf{h}_{k}^{H}\mathbf{p}_{k}|^{2}}{\sigma_{p,k}} \geq \alpha_{k},  
		\ \frac{|\mathbf{h}_{k}^{H}\mathbf{p}_{k}|^{2}}{\sigma_{p,k}} \leq \beta_{k},  
		\ \frac{|\mathbf{h}_{k}^{H}\mathbf{p}_{c}|^{2}}{\sigma_{c,k}} \geq \omega_{k},  \label{eqn:alphabetaomega}\\ 
		& \quad\    \sigma_{p,k} \hspace{-0.1cm}\geq\hspace{-0.1cm} 1\hspace{-0.1cm}+\hspace{-0.4cm}\sum_{i \in \mathcal{K}, i \neq k}\hspace{-0.3cm}|\mathbf{h}_{k}^{H}\mathbf{p}_{i}|^{2},  \sigma_{c,k} \hspace{-0.1cm}\geq\hspace{-0.1cm} 1\hspace{-0.1cm}+\hspace{-0.1cm}\sum_{i \in \mathcal{K}}|\mathbf{h}_{k}^{H}\mathbf{p}_{i}|^{2},   \forall k \in \mathcal{K} \label{eqn:sigma_final}, \\
		& \quad\  0 \leq z_{k} \leq 1, \quad  \forall k \in \mathcal{K} \label{eqn:z_final}, \\
		& \quad  \prod_{k \in \mathcal{K}}z_{k}= 0,  \label{eqn:prodz_prefinal} \\
		& \quad  \eqref{eqn:common_rate_2}, \eqref{eqn:transmit_power}. \nonumber 
	\end{alignat}
	\label{eqn:formulation_prefinal}
\end{subequations}
\vspace{-0.5cm}

\hspace{-0.25cm} We also add the new constraint \eqref{eqn:prodz_prefinal} for easy convergence of the proposed solution after the binary constraint is relaxed. We apply first-order Taylor approximation to relax the constraints \eqref{eqn:bigM2_prefinal}, \eqref{eqn:alphabetaomega}, \eqref{eqn:prodz_prefinal} and obtain the following convex problem.
\begin{subequations}
	\begin{alignat}{3}
	&\min_{\substack{\mathbf{c}, \mathbf{P}, \mathbf{z}, \boldsymbol{\alpha}, \\ \boldsymbol{\beta}, \boldsymbol{\omega}, 	\boldsymbol{\sigma}_{p}, \boldsymbol{\sigma}_{c}}} \sum_{k\in\mathcal{K}}(t+1-V_{t+1,k})g(X_{t+1,k},\widehat{X}_{t+1,k})z_{k}  \\
	&\ \ \text{s.t.}  \   c_{k}+\frac{\log(1+\beta^{[n-1]}_{k})}{\log(2)}+\frac{(\beta_{k}-\beta^{[n-1]}_{k})}{(1+\beta^{[n-1]}_{k})\log(2)}\nonumber\\
	&  \quad\quad\quad\quad\quad\quad\quad\quad\quad -I_{k} \leq M(1-z_{k}), \quad  \forall k \in \mathcal{K}, \label{eqn:bigM2_final} \\
	& \quad  \frac{2\mathcal{R}\{(\mathbf{p}^{[n-1]}_{k})^{H}\mathbf{h}_{k}\mathbf{h}_{k}^{H}\mathbf{p}_{k}\}}{\sigma^{[n-1]}_{p,k}}-\frac{|\mathbf{h}_{k}^{H}\mathbf{p}^{[n-1]}_{k}|^{2}}{(\sigma^{[n-1]}_{p,k})^{2}}\sigma_{p,k} \geq \alpha_{k},  \label{eqn:alpha_final}\\ 
	& \quad  \frac{2\mathcal{R}\{(\mathbf{p}^{[n-1]}_{k})^{H}\mathbf{h}_{k}\mathbf{h}_{k}^{H}\mathbf{p}_{k}\}}{\sigma^{[n-1]}_{p,k}}-\frac{|\mathbf{h}_{k}^{H}\mathbf{p}^{[n-1]}_{k}|^{2}}{(\sigma^{[n-1]}_{p,k})^{2}}\sigma_{p,k} \leq \beta_{k},  \label{eqn:beta_final}\\ 
	& \quad  \frac{2\mathcal{R}\{(\mathbf{p}^{[n-1]}_{c})^{H}\mathbf{h}_{k}\mathbf{h}_{k}^{H}\mathbf{p}_{c}\}}{\sigma^{[n-1]}_{c,k}}-\frac{|\mathbf{h}_{k}^{H}\mathbf{p}^{[n-1]}_{c}|^{2}}{(\sigma^{[n-1]}_{c,k})^{2}}\sigma_{c,k} \geq \omega_{k},  \label{eqn:alpha_final}\\ 
	& \quad   \prod_{k \in \mathcal{K}}z^{[n-1]}_{k}+\sum_{k \in \mathcal{K}}(z_{k}-z^{[n-1]}_{k})\prod_{\substack{l \in \mathcal{K}, \ \ l \neq k}}z^{[n-1]}_{l} = 0,  \label{eqn:prodz_final} \\
	& \quad\quad\    \eqref{eqn:bigM1_prefinal}, \eqref{eqn:common_rate_1_prefinal}, \eqref{eqn:sigma_final},  \eqref{eqn:z_final}, \eqref{eqn:common_rate_2}, \eqref{eqn:transmit_power}. \nonumber 
	\end{alignat}
	\label{eqn:formulation_final}
\end{subequations}
\vspace{-0.5cm}

\begin{figure}[t!]
 \removelatexerror
  \begin{algorithm}[H]
	 \caption{Proposed Algorithm}
		\label{alg:algorithm}
			$n \gets 1$, $\mathbf{P}^{[0]}$, $\mathbf{z}^{[0]}$, $\boldsymbol{\beta}^{[0]}$, $\boldsymbol{\sigma}^{[0]}_{c}$, $\boldsymbol{\sigma}^{[0]}_{p}$, $\sum\Delta_{\mathrm{AoII},k}^{[0]}$ \\
			\While{$|\sum\Delta_{\mathrm{AoII},k}^{[n]}-\sum\Delta_{\mathrm{AoII},k}^{[n-1]}| \geq \epsilon$ }{
			Solve \eqref{eqn:formulation_final} using interior-point methods to obtain $\sum\Delta_{\mathrm{AoII},k}^{*}, \mathbf{P}^{*}, \mathbf{z}^{*}, \boldsymbol{\alpha}^{*},\boldsymbol{\beta}^{*}, \boldsymbol{\omega}^{*}, \boldsymbol{\sigma}^{*}_{c}, \boldsymbol{\sigma}^{*}_{p}$; \\
			$\sum\Delta_{\mathrm{AoII},k}^{[n]} \gets \Delta_{\mathrm{AoII},k}^{*}$, $\mathbf{P}^{[n]} \gets \mathbf{P}^{*}$, $\mathbf{z}^{[n]} \gets \mathbf{z}^{*}$, $\boldsymbol{\beta}^{[n]} \gets \boldsymbol{\beta}^{*}$, $\boldsymbol{\sigma}_{c}^{[n]} \gets \boldsymbol{\sigma}_{c}^{*}$, $\boldsymbol{\sigma}_{p}^{[n]} \gets \boldsymbol{\sigma}_{p}^{*}$; \\
			$n \gets n + 1$.			
			}
			\Return $\mathbf{P} \gets \mathbf{P}^{[n]}$, $\mathbf{z} \gets \mathbf{z}^{[n]}$
  \end{algorithm}
  \vspace{-0.7cm}
\end{figure}

The parameters with superscript $[n-1]$ represent the variables that the Taylor approximation is applied around. We propose an iterative algorithm to obtain the precoder matrix and user scheduling, which is given in Algorithm~\ref{alg:algorithm}. At iteration $n$, the variables obtained at iteration $n-1$
are used in the related constraints to solve the convex problem using interior-point methods \cite{boyd2004}. The iterations continue until the objective function converges within a tolerance of $\epsilon$. We note that the algorithm output $\mathbf{z}$ contains the relaxed user scheduling indicators. Therefore, we apply the following method to obtain binary scheduling indicators.  
\begin{align}
	z^{\prime}_{k}=
	\begin{cases}
    1, \ \text{if } z_{k}=1 \ \text{or} \ c_{k}(\mathbf{P})+R_{k}(\mathbf{P}) \leq I_{k} , \\
    0, \ \text{otherwise.}
\end{cases}
\label{eqn:z}
\end{align}
The rates in \eqref{eqn:z} are calculated using the precoders obtained from Algorithm~1. The convergence of Algorithm~\ref{alg:algorithm} is guaranteed since the solution of \eqref{eqn:formulation_final} at iteration $n$ is also a feasible solution at iteration $n+1$. Therefore, the objective function is monotonically decreasing and is bounded below by zero. The worst case complexity of solving \eqref{eqn:formulation_final} using interior-point methods is $\mathcal{O}((NK)^{3.5}\log(1/\epsilon))$.
  
\section{Numerical Results}
We analyse the performance of the proposed algorithm and compare the performance of RSMA with that of SDMA by numerical results. SDMA is special case of RSMA when we do not allocate any power to the common stream $s_{c}$ and encode $W_{k}$ into $s_{k}$. We use CVX toolbox in MATLAB to solve \eqref{eqn:formulation_final} in Algorithm~\ref{alg:algorithm} \cite{cvx2008}. We use $\epsilon=10^{-4}$, $M=20$, and $I_{k}=I$, $\forall k \in \mathcal{K}$, throughout all simulations. The random process $X_{t,k}$ and $\widehat{X}_{t,k}$ are modelled as a discrete-time white Gaussian noise with unit variance and independent for all users \footnote{We note that when a user is scheduled for transmission, {i.e}, $c_{k}+R_{k} \geq I_{k}$, its update $X_{t,k}$ is successfully transmitted (and successfully decoded) as mentioned in Section III, however, the transceiver algorithms that are used to achieve this rate, such as modulation and coding schemes, equalizer and decoder architectures are out of the scope of this work.}. 
\begin{figure}[t!]
	\begin{subfigure}{.24\textwidth}
		\centerline{\includegraphics[width=1.8in,height=1.8in,keepaspectratio]{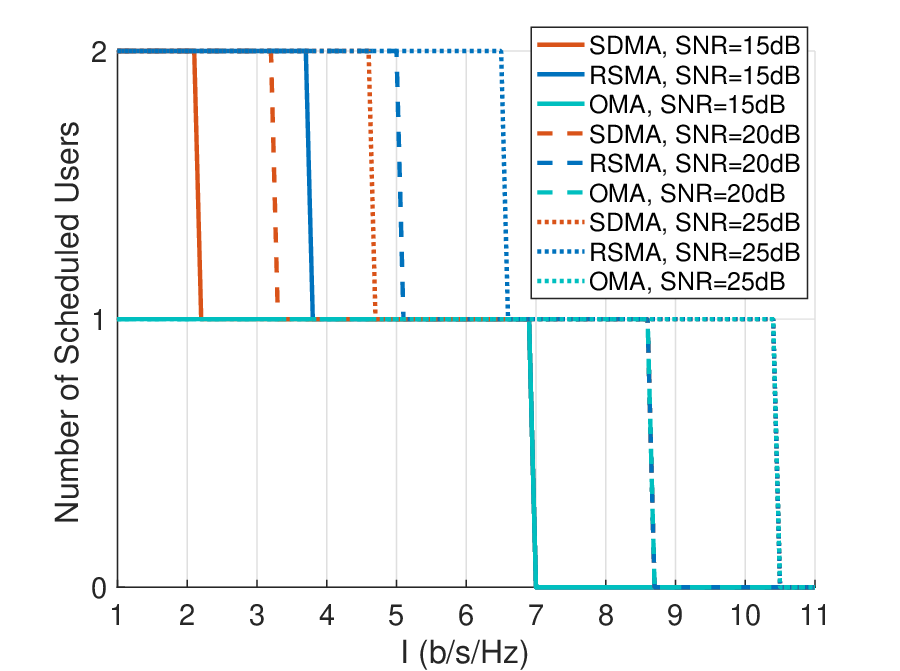}}
		\vspace{-0.2cm}
		\caption{$\theta=\pi/18$.}
		\label{fig:N4K2_GeoChan_Angle10_NumSch}
	\end{subfigure}
	\begin{subfigure}{.24\textwidth}
		\centerline{\includegraphics[width=1.8in,height=1.8in,keepaspectratio]{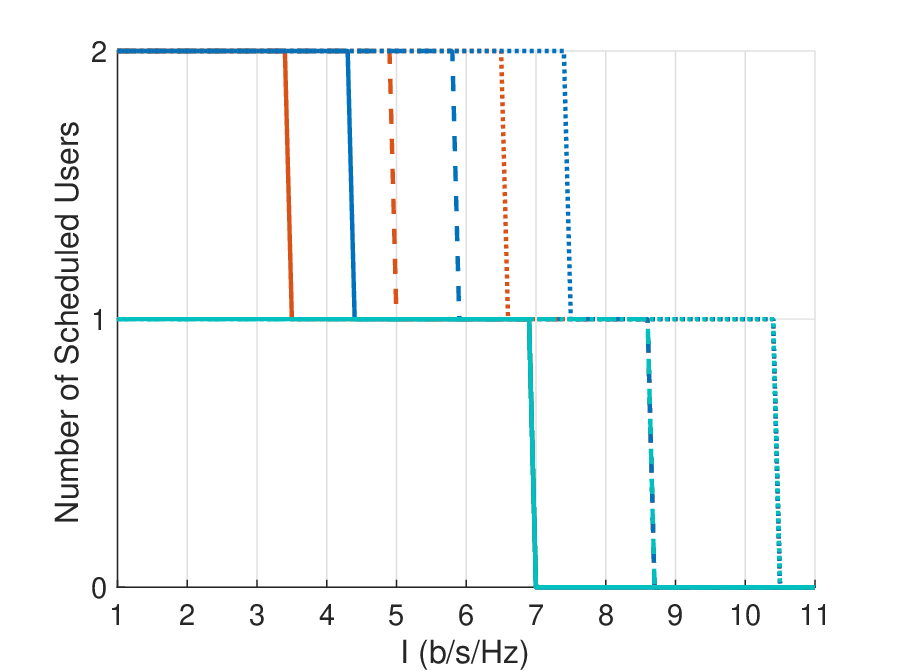}}
		\vspace{-0.2cm}
		\caption{$\theta=\pi/9$.}
		\label{fig:N4K2_GeoChan_Angle20_NumSch}
	\end{subfigure}
	\newline
	\begin{subfigure}{.24\textwidth}
		\centerline{\includegraphics[width=1.8in,height=1.8in,keepaspectratio]{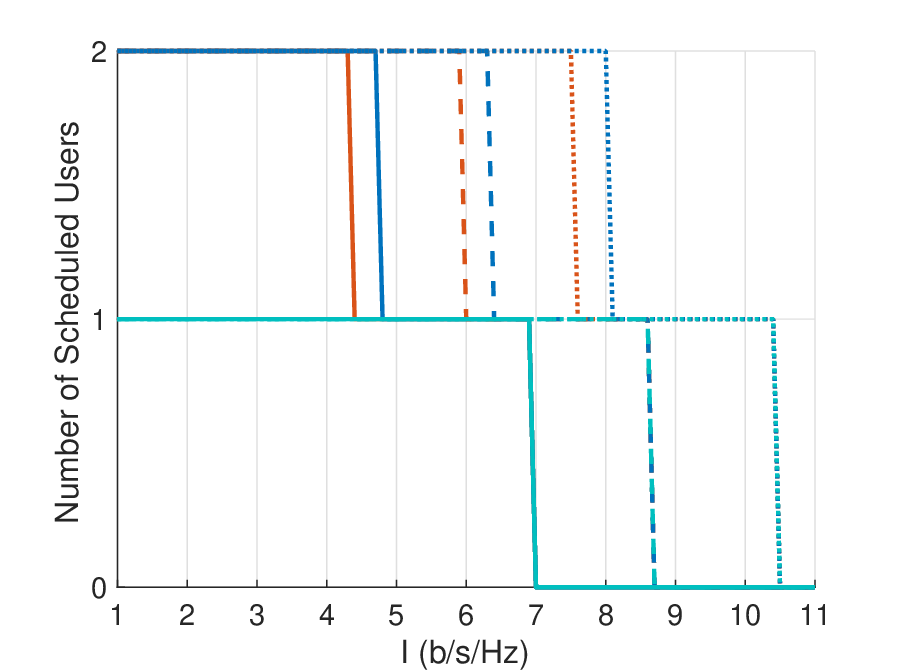}}
		\vspace{-0.2cm}
		\caption{$\theta=\pi/6$.}
		\label{fig:N4K2_GeoChan_Angle30_NumSch}
	\end{subfigure}
	\begin{subfigure}{.24\textwidth}
		\centerline{\includegraphics[width=1.8in,height=1.8in,keepaspectratio]{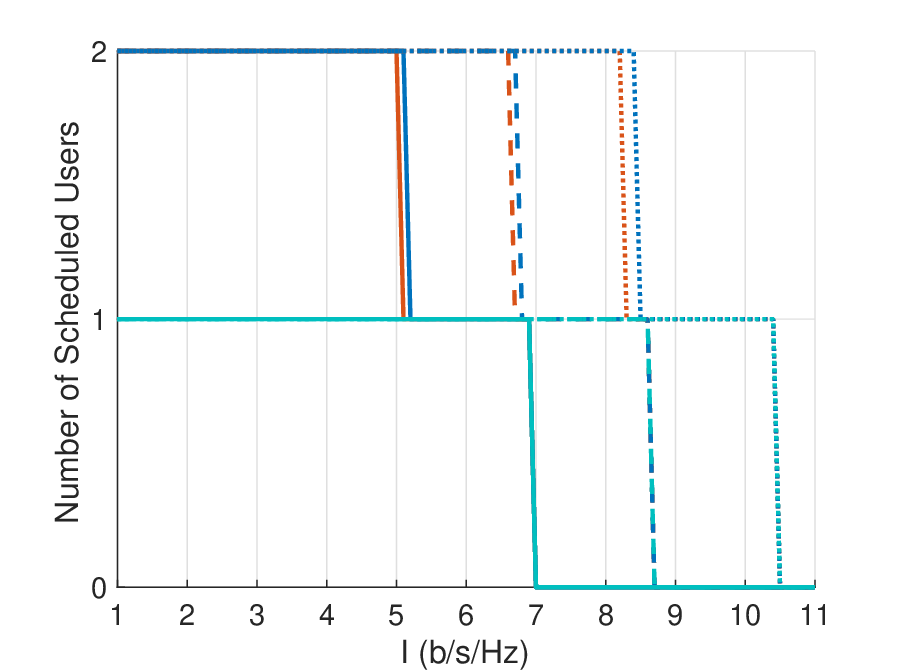}}
		\vspace{-0.2cm}
		\caption{$\theta=2\pi/9$.}
		\label{fig:N4K2_GeoChan_Angle40_NumSch}
	\end{subfigure}
	\vspace{-0.5cm}
	\caption{Number of scheduled users vs. $I$.} 
	\vspace{-0.5cm}
	\label{fig:N4K2_GeoChan_NumSch}
\end{figure}

We start investigating the performance of the proposed design using a simple channel model in a two-user system with $N=4$, $V_{t}=[0,0]$ and $t=1$. Let us define the channels $\mathbf{h}_{1}=[1\ 1\ 1\ 1]^{H}$ and $\mathbf{h}_{2}=[1\ e^{j\theta}\ e^{j2\theta}\ e^{j3\theta}]^{H}$, where $\theta$ represents the angle between the channels of the users \cite{mao_2018}.  Fig.~\ref{fig:N4K2_GeoChan_NumSch} shows the number of scheduled users using RSMA, SDMA, and Orthogonal Multiple Access (OMA) with respect to $I$ and and for various SNR values. We note that OMA can schedule only one user in a given time slot, {\sl i.e.}, between $t$ and $t+1$. Accordingly, RSMA and SDMA boil down to OMA when they can schedule one user. 

As seen from Fig.~\ref{fig:N4K2_GeoChan_NumSch}, RSMA can schedule both users for larger $I$ values than SDMA, especially in scenarios where the users are spatially close. In such scenarios, cancelling the multi-user interference becomes challenging, which affects the SINR of the interference free streams, and thus limits the achievable rates. This results in SDMA failing to satisfy the rate requirements of one of the users when $I$ is increased above certain levels.
RSMA achieves an enhanced performance by splitting the required rate between common and private streams and increasing the achievable rate of each user using the common stream when the using the private streams only is not enough to satisfy the rate requirements. The results also show that RSMA can ensure better user fairness than SDMA in the considered framework as it can schedule more users than SDMA for larger values of $I$. Both schemes boil down to OMA starting from certain $I$ values (SDMA boils down to OMA for smaller values of $I$ than those of RSMA) and until $I$ is smaller than capacity of the channel, after which none of the schemes can schedule any users.

\begin{figure}[t!]
	\begin{subfigure}{.24\textwidth}
		\centerline{\includegraphics[width=1.8in,height=1.8in,keepaspectratio]{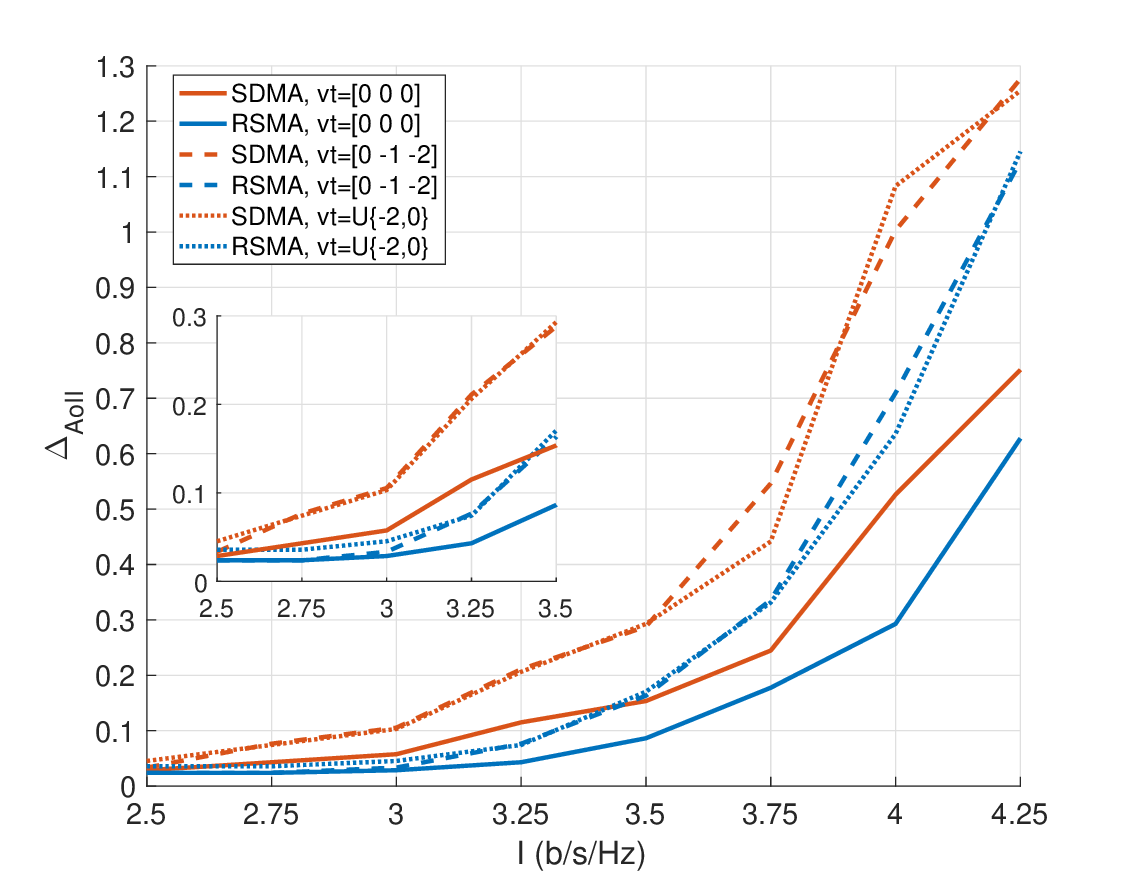}}
		\vspace{-0.2cm}
		\caption{SNR=$15$dB.}
		\label{fig:AoII_N4K3_Rayleigh_15dB}
	\end{subfigure}
	\begin{subfigure}{.24\textwidth}
		\centerline{\includegraphics[width=1.8in,height=1.8in,keepaspectratio]{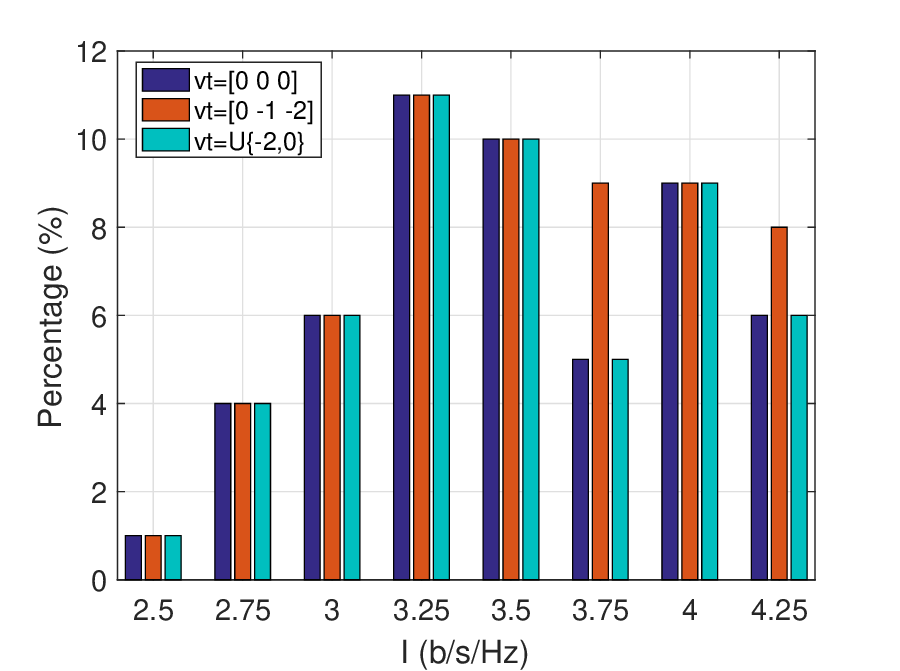}}
		\vspace{-0.2cm}
		\caption{SNR=$15$dB.}
		\label{fig:Percentage_N4K3_Rayleigh_15dB}
	\end{subfigure}
	\newline
	\begin{subfigure}{.24\textwidth}
		\centerline{\includegraphics[width=1.8in,height=1.8in,keepaspectratio]{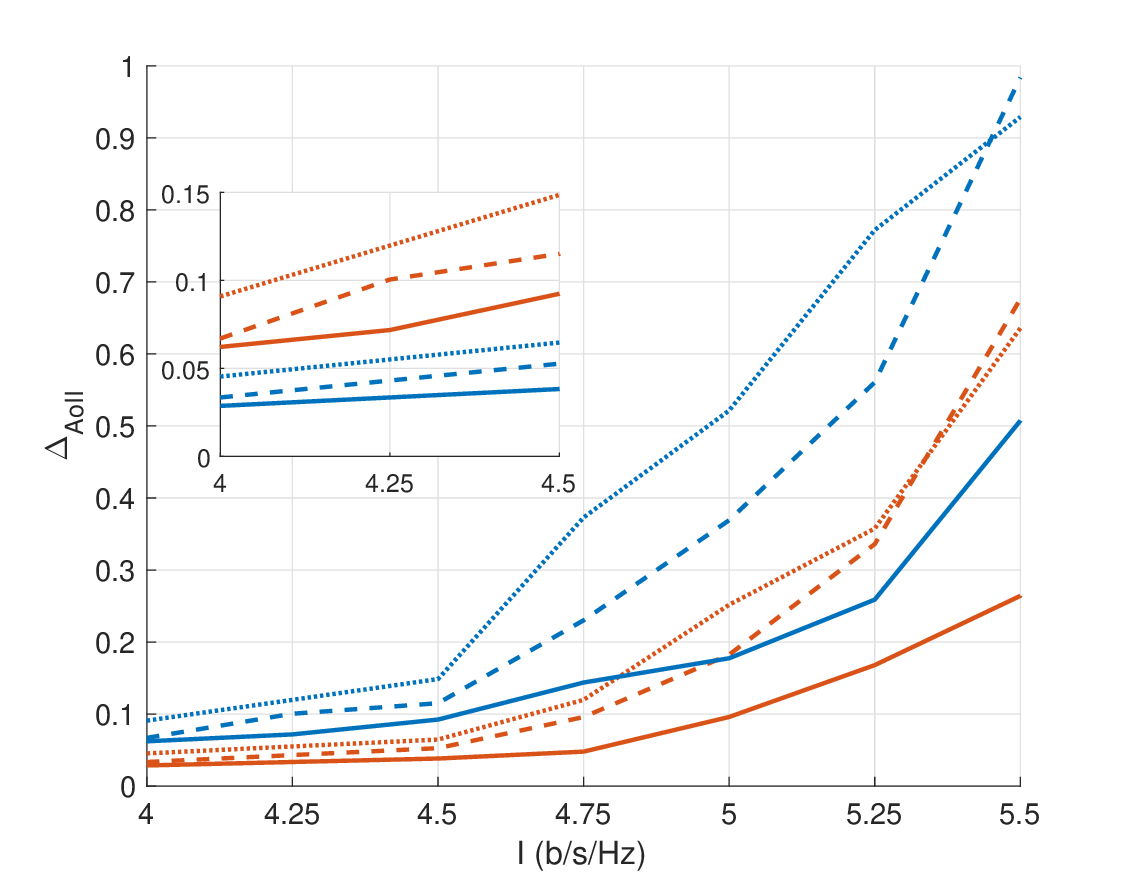}}
		\vspace{-0.2cm}
		\caption{SNR=$20$dB.}
		\label{fig:AoII_N4K3_Rayleigh_20dB}
	\end{subfigure}
	\begin{subfigure}{.24\textwidth}
		\centerline{\includegraphics[width=1.8in,height=1.8in,keepaspectratio]{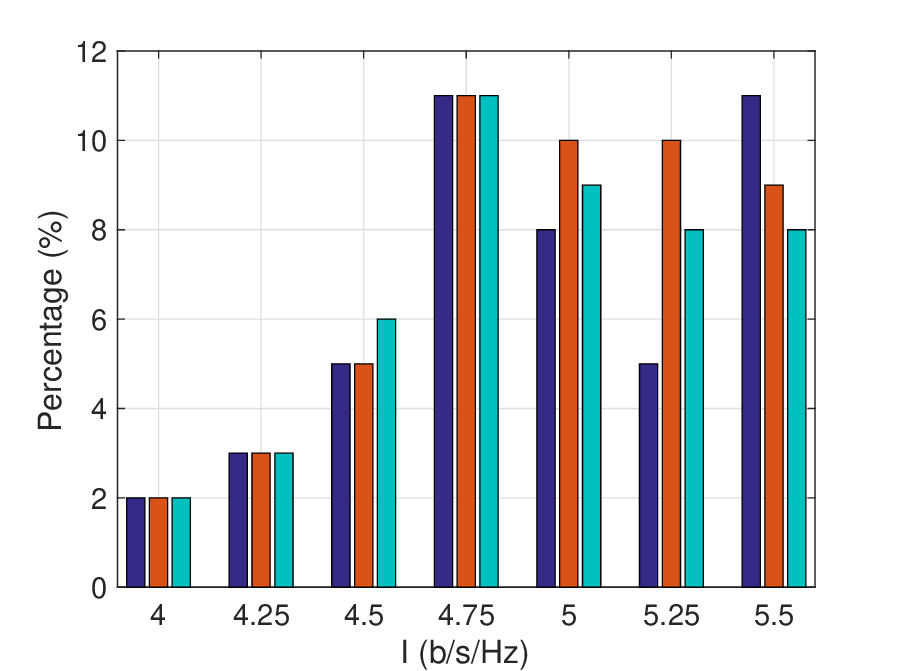}}
		\vspace{-0.2cm}
		\caption{SNR=$20$dB.}
		\label{fig:Percentage_N4K3_Rayleigh_20dB}
	\end{subfigure}
	\vspace{-0.5cm}
	\caption{AoII and scheduling performance, $N=4$, $K=3$. }
	\vspace{-0.3cm}
	\label{fig:N4K3_Rayleigh}
\end{figure}

\begin{figure}[t!]
	\begin{subfigure}{.24\textwidth}	\centerline{\includegraphics[width=1.9in,height=1.9in,keepaspectratio]{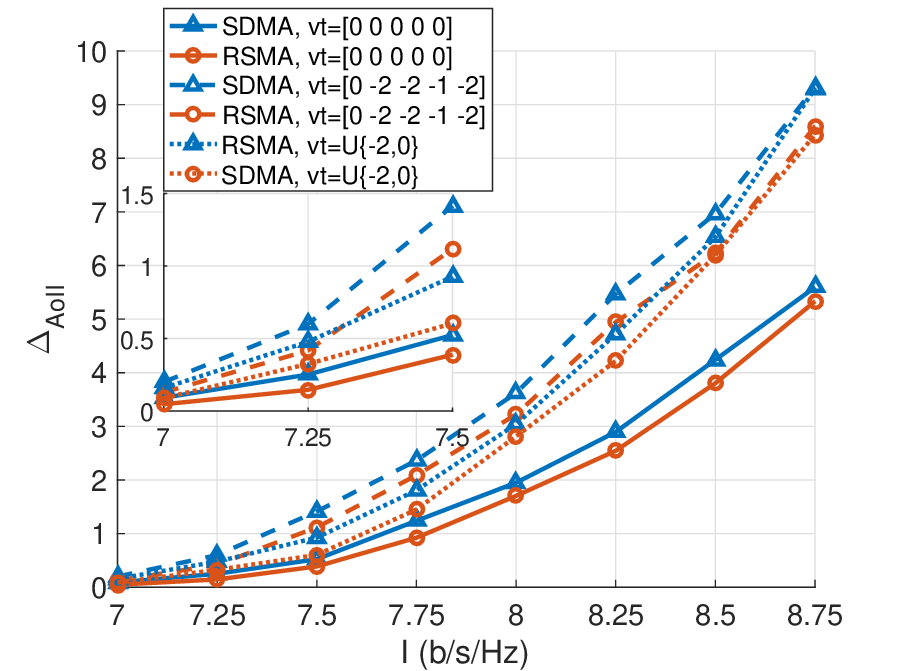}}
		\vspace{-0.2cm}
		\caption{SNR=$25$dB.}
		\label{fig:AoII_N8K5_Rayleigh_25dB}
	\end{subfigure}
	\begin{subfigure}{.24\textwidth}
	\centerline{\includegraphics[width=1.9in,height=1.9in,keepaspectratio]{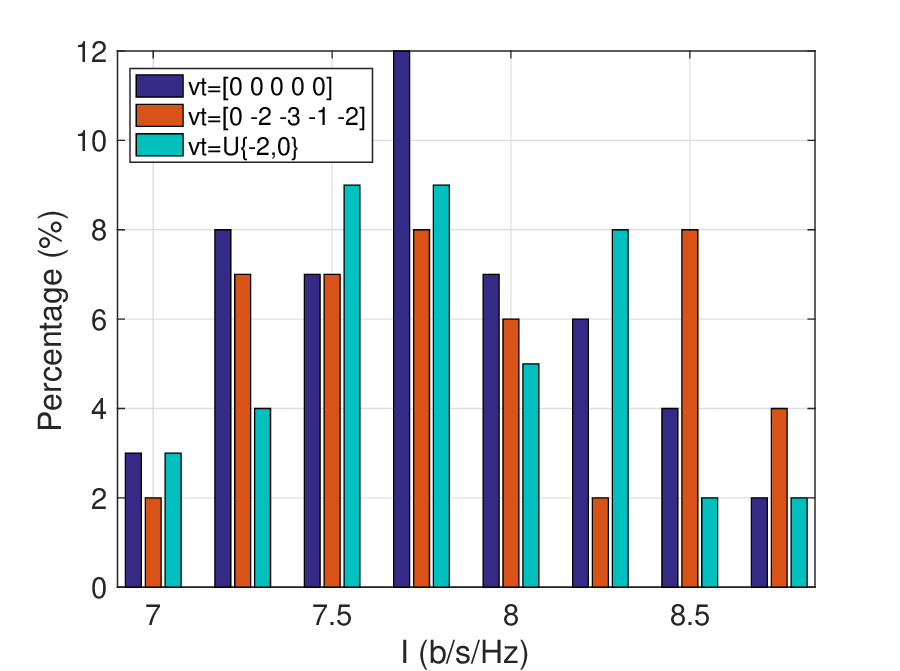}}
		\vspace{-0.2cm}
		\caption{SNR=$25$dB.}
		\label{fig:Percentage_N8K5_Rayleigh_25dB}
	\end{subfigure}
	\newline
	\begin{subfigure}{.24\textwidth}
	\centerline{\includegraphics[width=1.9in,height=1.9in,keepaspectratio]{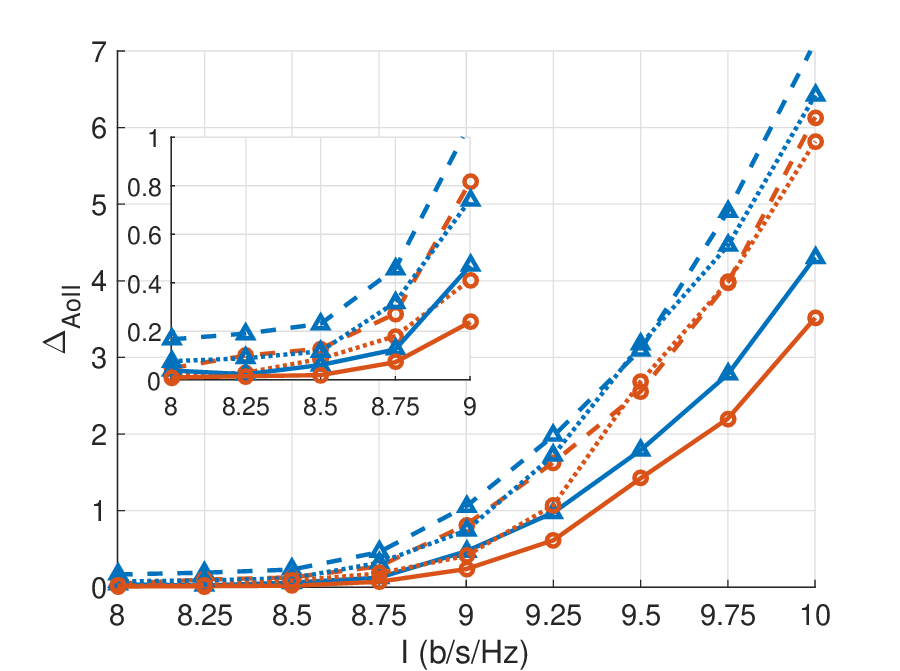}}
		\vspace{-0.2cm}
		\caption{SNR=$30$dB.}
		\label{fig:AoII_N8K5_Rayleigh_30dB}
	\end{subfigure}
	\begin{subfigure}{.24\textwidth}
		\centerline{\includegraphics[width=1.9in,height=1.9in,keepaspectratio]{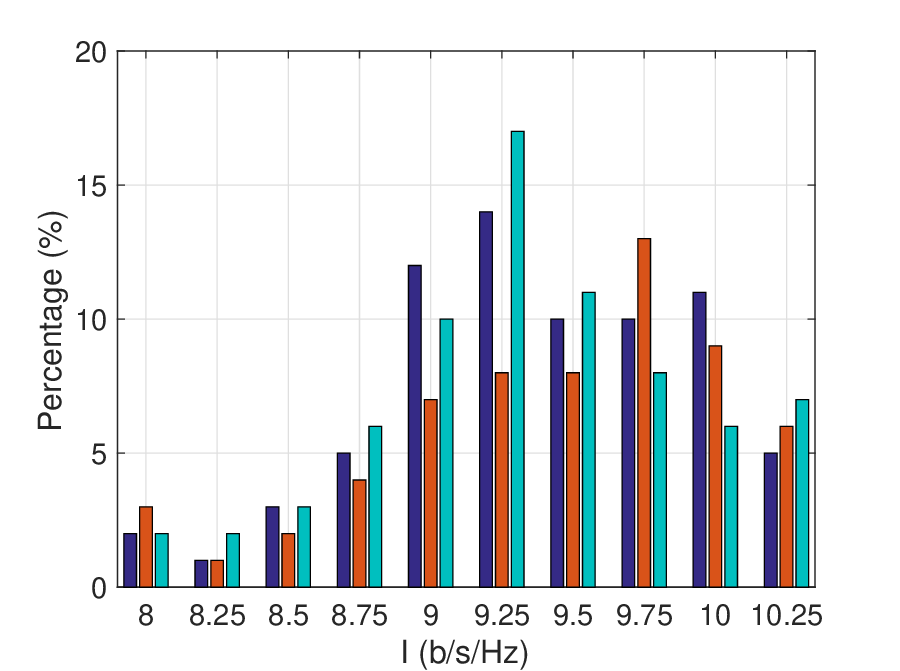}}
		\vspace{-0.2cm}
		\caption{SNR=$30$dB.}
		\label{fig:Percentage_N8K5_Rayleigh_30dB}
	\end{subfigure}
	\vspace{-0.5cm}
	\caption{AoII and scheduling performance, $N=8$, $K=5$. }
	\vspace{-0.7cm}
	\label{fig:N8K5_Rayleigh}
\end{figure}

Next, we investigate the performance of the system under Rayleigh fading channel. In Fig.~\ref{fig:N4K3_Rayleigh} and \ref{fig:N8K5_Rayleigh}, we demonstrate the average AoII with average taken over $100$ channel realizations for various $N$, $K$, $I$, and SNR values.  We also investigate the percentage of number of users scheduled using RSMA and SDMA by comparing their performance. For this purpose, we give the percentage of cases where RSMA schedules more users than SDMA in a time slot. We run the simulations for the cases where $V_{t,k}=0$, $\forall k \in \mathcal{K}$, $V_{t,k}$ are pre-defined values for all Monte-Carlo runs, and $V_{t,k}$ is chosen from a discrete uniform distribution at each Monte-Carlo run. As it is seen from the figures, RSMA achieves a lower AoII than SDMA for all considered scenarios. The results show that RSMA can schedule more users than SDMA for the same $I$ values and this achieves a lower average AoII.

\begin{figure}[t!]
	\begin{subfigure}{.24\textwidth}
		\centerline{\includegraphics[width=1.8in,height=1.8in,keepaspectratio]{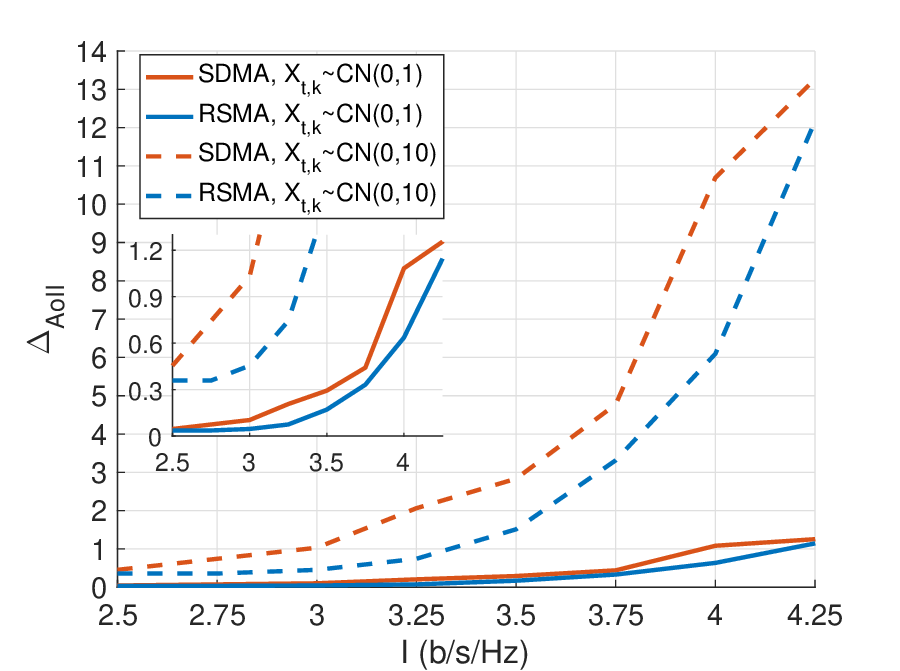}}
		\vspace{-0.2cm}
		\caption{AoII.}
		\label{fig:AoII_N4K3_Rayleigh_15dB_DiffProcess}
	\end{subfigure}
	\begin{subfigure}{.24\textwidth}
		\centerline{\includegraphics[width=1.8in,height=1.8in,keepaspectratio]{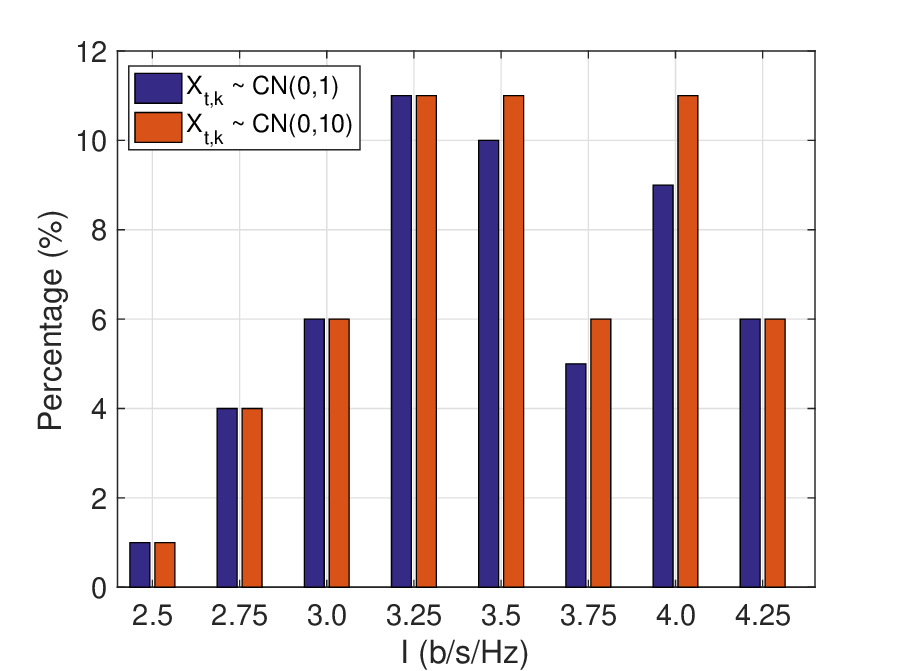}}
		\vspace{-0.2cm}
		\caption{Percentage.}
		\label{fig:Percentage_N4K3_Rayleigh_15dB_DiffProcess}
	\end{subfigure}
	\caption{AoII and scheduling performance for two different processes, $N=4$, $K=3$, SNR=$15$dB. }
	\vspace{-0.5cm}
	\label{fig:N4K3_Rayleigh_DiffProcess}
\end{figure}

Finally, we investigate the impact of information process distribution on the error performance. Fig.~\ref{fig:N4K3_Rayleigh_DiffProcess} gives the performance comparison of RSMA and SDMA when two different process distributions, $X_{t,k} \sim \mathcal{CN}(0,1)$ and $X_{t,k} \sim \mathcal{CN}(0,10)$, are considered. As seen from Fig.~\ref{fig:Percentage_N4K3_Rayleigh_15dB_DiffProcess}, the percentage of channel realizations where more users are allocated by RSMA is approximately the same for both process distributions. However, a significant difference is observed in the AoII values and AoII gain achieved by RSMA from Fig.~\ref{fig:AoII_N4K3_Rayleigh_15dB_DiffProcess} when a process with higher variance is considered. This is expected as a process having a distribution with higher variance may observe larger changes in value between time slots. We note here that the obtained results with a different process may change when a different $g(X_{t,k}, \widehat{X}_{t,k})$ is used, such as, $g_{\mathrm{thr}}(X_{t,k}, \widehat{X}_{t,k})=\mathds{1}\{|X_{t,k}-\widehat{X}_{t,k}|\geq c\}$.

The results presented in this section show that RSMA can schedule more users and achieve lower AoII depending on the channel conditions even under perfect CSIT. Accordingly, one would expect that the performance gain achieved by RSMA over SDMA in the considered framework is going to be even more significant under imperfect CSIT due to the enhanced performance of RSMA under multi-user interference.

\section{Conclusion}
In this work, we study the performance of RSMA for downlink multi-user communications in semantic-aware networks. We formulate a sum-AoII minimization problem to find the optimal user scheduling, precoding, and power allocation jointly. We apply big-M and SCA to the formulated problem and propose an iterative algorithm to solve it. By numerical results, we show that RSMA can achieve a lower AoII than SDMA under perfect CSIT.
Future work includes taking into account the effects imperfect CSIT, imperfect CSIR and SIC, hardware constraints, and scalability constraints in system design for practical applications and demonstrating the developed framework as a Proof-of-Concept.


\end{document}